\documentclass[12pt,superscriptaddress,aps,prd,preprint,showpacs]{revtex4}
\usepackage{amsmath, amsthm, amssymb, amsfonts, mathrsfs}
\usepackage{physics, braket, slashed, siunitx, bm}
\usepackage{bbold}
\usepackage{listings}
\usepackage{color}
\usepackage[usenames,dvipsnames]{xcolor}
\usepackage{graphicx}
\usepackage{float}
\usepackage{placeins}
\usepackage{booktabs}
\usepackage{hyperref}
\usepackage{soul}
\usepackage{ulem}

\hypersetup{
	unicode=false,          
	pdftoolbar=true,        
	pdfmenubar=true,        
	pdffitwindow=false,     
	pdfauthor={William},     
	colorlinks=true,       
	linkcolor=blue,          
	citecolor=red,        
	urlcolor=blue
}

\newcommand{\bea}{\begin{eqnarray}}
	\newcommand{\eea}{\end{eqnarray}}

\begin{document}
	
	\title{Lorentz-breaking Rarita-Schwinger model}
	
	\author{M. Gomes}
	\email{mgomes@if.usp.br}
\affiliation{Instituto de F\'\i sica, Universidade de S\~ao Paulo\\
Caixa Postal 66318, 05315-970, S\~ao Paulo, SP, Brazil}
	
	\author{T. Mariz}
	\email{tmariz@fis.ufal.br}
\affiliation{Instituto de F\'\i sica, Universidade Federal de Alagoas, 57072-900, Macei\'o, AL, Brazil}

	\author{J. R. Nascimento}
	\email{jroberto@fisica.ufpb.br}
	\affiliation{Departamento de Física, Universidade Federal da Paraíba,~\\
		Caixa Postal 5008, 58051-970, João Pessoa, Paraíba, Brazil}
	
	\author{A. Yu. Petrov}
	\email{petrov@fisica.ufpb.br}
	\affiliation{Departamento de Física, Universidade Federal da Paraíba,~\\
		Caixa Postal 5008, 58051-970, João Pessoa, Paraíba, Brazil}
	\date{\today}
	\begin{abstract}
		In this paper, we formulate the Lorentz-breaking extension for the spin-3/2 field theory and couple it to the Abelian gauge field in a Lorentz violating (LV) manner. Next, we calculate the lower LV quantum corrections, that is, the Carroll-Field-Jackiw (CFJ) term, which, being superficially divergent, turns out to be finite but ambiguous, and also the higher-derivative CFJ term. Besides, we compute the aether term, being the lowest CPT-even LV term, proportional to the second order in the LV vector.
	\end{abstract}
	\maketitle

The possibility of Lorentz symmetry breaking is actively discussed now, including not only theoretical but also experimental studies \cite{datatables}. Various LV extensions have been discussed within the context of the Lorentz-violating Standard Model Extension (LV SME) originally formulated in \cite{ColKost1,ColKost2} and further generalized in \cite{KosLi}. Essentially, most of them represent themselves as extensions of spinor QED or its non-Abelian analogues constructed by adding some LV operators. The LV couplings of gauge fields to scalar ones have been considered within the Higgs sector of LV SME (see, e.g., \cite{ColKost1,ColKost2}). Further, various LV extensions of gravity, that is, extensions of LV SME to curved space-time, have been considered \cite{KosGra,KosLiGrav}. At the same time, it is interesting to study other LV theories that have not been included in LV SME. In this paper, we follow this idea, introducing a LV generalization of the Rarita-Schwinger (RS) theory \cite{RS}, describing the spin-3/2 field known to play an essential role within the supergravity context, where this field corresponds to the gravitino (for a review on supergravity, see, e.g., \cite{SGRS}). While the RS Lagrangian presents certain drawbacks implying the possibility of superluminal solutions \cite{Velo,Velo1} (we note nevertheless that the situation is improved in supergravity, see, e.g., \cite{DF}), in a certain sense, we expect our study to be a prototype for further studies of LV extensions of supergravity. Within this paper, we couple the spin-3/2 field to the electromagnetic field, introduce the simplest LV term, for the first time within studies of the gravitino, and obtain the lower LV quantum corrections, that is, the Carroll-Field-Jackiw (CFJ) term \cite{CFJ}, and then, the higher-derivative CFJ-like \cite{MWS} and the aether ones \cite{aetherCarroll}.

The most general spin-3/2 free Lagrangian looks like \cite{Hab} (see also, e.g., \cite{Pascal}, where issues related to canonical quantization and phenomenological aspects of this theory were discussed, and references therein):
\bea
\label{spin32}
{\cal L}=\bar{\psi}_{\mu}\Lambda^{\mu\nu}\psi_{\nu},
\eea
where, in the momentum space, 
\bea
\Lambda^{\mu\nu}&=&(\slashed{p}-M)\eta^{\mu\nu}
+A(\gamma^{\mu}p^{\nu}+\gamma^{\nu}p^{\mu})+\frac{1}{2}(3A^2+2A+1)\gamma^{\mu}\slashed{p}\gamma^{\nu}+\nonumber\\ &+&M(3A^2+3A+1)\gamma^{\mu}\gamma^{\nu}.
\eea
Here $A$ is a real number. The usual Rarita-Schwinger case \cite{RS} corresponds to the requirement $A=-1$, which will be adopted henceforth, so that one has
\bea
\Lambda^{\mu\nu}=(\slashed{p}-M)\eta^{\mu\nu}
-(\gamma^{\mu}p^{\nu}+\gamma^{\nu}p^{\mu})+\gamma^{\mu}\slashed{p}\gamma^{\nu}+M\gamma^{\mu}\gamma^{\nu}.
\eea
Then, the corresponding propagator is
\bea
G^{\mu\nu}(p)=(\Lambda^{-1})^{\mu\nu}=\frac{\slashed{p}+M}{p^2-M^2}(\eta^{\mu\nu}-\frac{1}{3}\gamma^{\mu}\gamma^{\nu}-\frac{2p^{\mu}p^{\nu}}{3M^2}-\frac{\gamma^{\mu}p^{\nu}-\gamma^{\nu}p^{\mu}}{3M}).
\eea
We note that at $M=0$, $\Lambda^{\mu\nu}$ is transverse, i.e., the theory (\ref{spin32}) displays the gauge symmetry in this case, while the mass term proportional to $M$ breaks the gauge symmetry. Therefore, at $M=0$, to define the propagator consistently, one needs to introduce a gauge fixing term for the RS field. To avoid this problem, within our studies, we assume $M\neq 0$.

It is clear that the electromagnetic field can be coupled to the RS field through ``covariantizing" the action by promotion of the usual derivative to the gauge covariant one, $\partial_{\mu}\to D_{\mu}=\partial_{\mu}-ieA_{\mu}$ \cite{Pill}, or, as is the same in the momentum space, to replace $p_{\mu}\to p_{\mu}+eA_{\mu}$. By a whole analogy with the spinor QED, the resulting theory, at $A=-1$, is
\bea
\label{lagr}
{\cal L}=\bar{\psi}_{\mu}\left((i\slashed{D}-M)\eta^{\mu\nu}
-i(\gamma^{\mu}D^{\nu}+\gamma^{\nu}D^{\mu})+i\gamma^{\mu}\slashed{D}\gamma^{\nu}+M\gamma^{\mu}\gamma^{\nu}+\slashed{b}\gamma_5\eta^{\mu\nu}\right)\psi_{\nu}.
\eea
To break the Lorentz symmetry, we have introduced here the term proportional to the axial vector $b_{\mu}$. We note that this term will allow generating not only the CFJ term but also the aether term (which nevertheless will be obtained in our paper with the use of another coupling) and higher-derivative terms. 

The Lagrangian (\ref{lagr}) is gauge invariant under usual $U(1)$ transformations $A_{\mu}\to A_{\mu}-\partial_{\mu}\xi$, $\psi_{\mu}\to e^{i\xi}\psi_{\mu}$.  In this case, the RS field transforms as the usual matter. As we already noted, if we had $M=0$, the free Lagrangian of the RS field would possess its own "gauge" symmetry (indeed, in this case, it looks like $\epsilon^{\mu\nu\lambda\rho}\bar{\psi}_{\mu}\gamma_{\nu}\gamma_5\partial_{\lambda}\psi_{\rho}$, with gauge transformations $\psi_{\mu}\to\psi_{\mu}+\partial_{\mu}\varphi$, where $\varphi$ is completely distinct of the above parameter $\xi$ since their mass dimensions are different, 0 for $\xi$ and $1/2$ for $\varphi$, and while $\xi$ is a bosonic parameter, $\varphi$ is a fermionic one), but the mass term proportional to $M$ evidently breaks this symmetry. It is interesting to note, nevertheless, that the CFJ term does not depend on $M$, which allows us to expect that the results we obtain for this term will be valid in the massless case as well, but this is not so for higher-order corrections which depend on $M$ explicitly.

\begin{figure}[htbp]
  \includegraphics{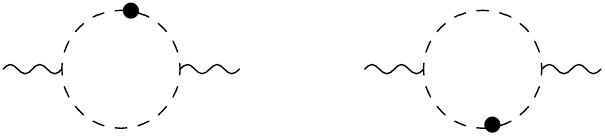}
\caption{CFJ-like contributions to the two-point function of the vector field.}
\label{fig:1}       
\end{figure}

We see that we have only triple vertices. Actually, they can be read off from
\bea
\label{vert}
V_e=e\bar{\psi}_{\mu}\left(\slashed{A}\eta^{\mu\nu}
-(\gamma^{\mu}A^{\nu}+\gamma^{\nu}A^{\mu})+\gamma^{\mu}\slashed{A}\gamma^{\nu}\right)\psi_{\nu}=e\bar{\psi}_{\mu}T^{\mu\nu}[A]\psi_{\nu}.
\eea
As a result, we have the graphs contributing to the CFJ term similar to the usual Lorentz-violating QED (see, e.g., \cite{ColKost2}), but with completely different vertices. Explicitly, our contributions have the forms given in Fig. \ref{fig:1}.

Their expressions are
\bea
\Pi_a&=&-e^2{\rm tr}\int\frac{d^4k}{(2\pi)^4}T^{\mu\nu}[A(-p)]G_{\nu\alpha}(k)\slashed{b}\gamma_5\eta^{\alpha\beta}G_{\beta\rho}(k)T^{\rho\sigma}[A(p)]G_{\sigma\mu}(k+p),\nonumber\\
\Pi_b&=&-e^2{\rm tr}\int\frac{d^4k}{(2\pi)^4}T^{\mu\nu}[A(-p)]G_{\nu\rho}(k)T^{\rho\sigma}[A(p)]G_{\sigma\alpha}(k+p)\slashed{b}\gamma_5\eta^{\alpha\beta}G_{\beta\mu}(k+p).
\eea
After writing the vertices described by $T^{\mu\nu}[A]$ factors, in the explicit form (\ref{vert}), we obtain
\bea
\label{twomin}
\Pi_a&=&-e^2{\rm tr}\int\frac{d^4k}{(2\pi)^4}A_{\lambda}(-p)A_{\tau}(p)
(\gamma^{\lambda}\eta^{\mu\nu}-\gamma^{\mu}\eta^{\nu\lambda}-\gamma^{\nu}\eta^{\mu\lambda}+\gamma^{\mu}\gamma^{\lambda}\gamma^{\nu})G_{\nu\alpha}(k)\slashed{b}\gamma_5\eta^{\alpha\beta}G_{\beta\rho}(k)\times\nonumber\\&\times&
(\gamma^{\tau}\eta^{\rho\sigma}-\gamma^{\rho}\eta^{\tau\sigma}-\gamma^{\sigma}\eta^{\tau\rho}+\gamma^{\rho}\gamma^{\tau}\gamma^{\sigma})G_{\sigma\mu}(k+p),\nonumber\\
\Pi_b&=&-e^2{\rm tr}\int\frac{d^4k}{(2\pi)^4}A_{\lambda}(-p)A_{\tau}(p)
(\gamma^{\lambda}\eta^{\mu\nu}-\gamma^{\mu}\eta^{\nu\lambda}-\gamma^{\nu}\eta^{\mu\lambda}+\gamma^{\mu}\gamma^{\lambda}\gamma^{\nu})G_{\nu\rho}(k)\times\nonumber\\&\times&
(\gamma^{\tau}\eta^{\rho\sigma}-\gamma^{\rho}\eta^{\tau\sigma}-\gamma^{\sigma}\eta^{\tau\rho}+\gamma^{\rho}\gamma^{\tau}\gamma^{\sigma})G_{\sigma\alpha}(k+p)\slashed{b}\gamma_5\eta^{\alpha\beta}G_{\beta\mu}(k+p).
\eea
To find the CFJ-like contribution (which is clearly nontrivial since this expression includes the $\gamma_5$ matrix), one can expand $G(k+p)$ in series in $p$ up to the first order:
\bea
G^{\mu\nu}(k+p)&=&G^{\mu\nu}(k)-\frac{1}{\slashed{k}-M}\slashed{p}\frac{1}{\slashed{k}-M}(\eta^{\mu\nu}-\frac{1}{3}\gamma^{\mu}\gamma^{\nu}-\frac{2k^{\mu}k^{\nu}}{3M^2}-\frac{\gamma^{\mu}k^{\nu}-\gamma^{\nu}k^{\mu}}{3M})-\nonumber\\ &-&
\frac{1}{\slashed{k}-M}
\left(\frac{2(k^{\mu}p^{\nu}+k^{\nu}p^{\mu})}{3M^2}+\frac{\gamma^{\mu}p^{\nu}-\gamma^{\nu}p^{\mu}}{3M}\right)+O(p^2).
\eea
Then, we can write
\bea
\label{defs}
\Pi_a=-e^2A_{\lambda}(-p)\Pi^{\lambda\tau}_a(p)A_{\tau}(p),\quad\, \Pi_b=-e^2A_{\lambda}(-p)\Pi^{\lambda\tau}_b(p)A_{\tau}(p),
\eea
so that $\Pi_b^{\lambda\tau}(-p)=\Pi_a^{\tau\lambda}(p)$ (an analogous relation takes place for two contributions to the CFJ self-energy tensor in the usual spinor QED with  one $\slashed{b}\gamma_5$ insertion, cf. \cite{ColKost2}). So, it is sufficient to find only $S_a$, i.e., to calculate $\Pi^{\lambda\tau}_a(p)$, given by
\bea
\label{Pia}
\Pi^{\lambda\tau}_a(p)&=&{\rm tr}\int\frac{d^4k}{(2\pi)^4}
(\gamma^{\lambda}\eta^{\mu\nu}-\gamma^{\mu}\eta^{\nu\lambda}-\gamma^{\nu}\eta^{\mu\lambda}+\gamma^{\mu}\gamma^{\lambda}\gamma^{\nu})G_{\nu\alpha}(k)\slashed{b}\gamma_5\eta^{\alpha\beta}G_{\beta\rho}(k)\times\nonumber\\&\times&
(\gamma^{\tau}\eta^{\rho\sigma}-\gamma^{\rho}\eta^{\tau\sigma}-\gamma^{\sigma}\eta^{\tau\rho}+\gamma^{\rho}\gamma^{\tau}\gamma^{\sigma})G_{\sigma\mu}(k+p)=\nonumber\\
&=&-{\rm tr}\int\frac{d^4k}{(2\pi)^4}
(\gamma^{\lambda}\eta^{\mu\nu}-\gamma^{\mu}\eta^{\nu\lambda}-\gamma^{\nu}\eta^{\mu\lambda}+\gamma^{\mu}\gamma^{\lambda}\gamma^{\nu})
\frac{\slashed{k}+M}{k^2-M^2}\times\nonumber\\&\times&
(\eta_{\nu\alpha}-\frac{1}{3}\gamma_{\nu}\gamma_{\alpha}-\frac{2k_{\nu}k_{\alpha}}{3M^2}-\frac{\gamma_{\nu}k_{\alpha}-\gamma_{\alpha}k_{\nu}}{3M})
\times\nonumber\\&\times&
\slashed{b}\gamma_5\eta^{\alpha\beta}\frac{\slashed{k}+M}{k^2-M^2}(\eta_{\beta\rho}-\frac{1}{3}\gamma_{\beta}\gamma_{\rho}-\frac{2k_{\rho}k_{\beta}}{3M^2}-\frac{\gamma_{\beta}k_{\rho}-\gamma_{\rho}k_{\beta}}{3M})\times\nonumber\\&\times&
(\gamma^{\tau}\eta^{\rho\sigma}-\gamma^{\rho}\eta^{\tau\sigma}-\gamma^{\sigma}\eta^{\tau\rho}+\gamma^{\rho}\gamma^{\tau}\gamma^{\sigma})
\times\nonumber\\&\times&\left(\frac{1}{\slashed{k}-M}\slashed{p}\frac{1}{\slashed{k}-M}(\eta_{\sigma\mu}-\frac{1}{3}\gamma_{\sigma}\gamma_{\mu}-\frac{2k_{\sigma}k_{\mu}}{3M^2}-\frac{\gamma_{\sigma}k_{\mu}-\gamma_{\mu}k_{\sigma}}{3M})+\right.\nonumber\\ &+&\left.
\frac{1}{\slashed{k}-M}
\left(\frac{2(k_{\sigma}p_{\mu}+k_{\mu}p_{\sigma})}{3M^2}+\frac{\gamma_{\sigma}p_{\mu}-\gamma_{\mu}p_{\sigma}}{3M}\right)
\right)+O(p^2).
\eea
It remains to find the trace and, afterward, the integral which badly (sextically) diverges. It is clear that $\Pi^{\lambda\tau}_a(p)\propto\epsilon^{\lambda\tau\alpha\beta}b_{\alpha}p_{\beta}$, so, our result yields the CFJ form.

In the first manner, we calculate all traces and integrals in $D=4-\epsilon$ dimensions. In this case, the corresponding contribution to the effective Lagrangian is
\bea
{\cal L}_{CFJ,1}=\frac{13e^2}{54\pi^2}\epsilon^{\lambda\mu\nu\rho}b_{\lambda}A_{\mu}\partial_{\nu}A_{\rho}.
\eea
Alternatively, we can calculate all traces in (\ref{Pia}) in four dimensions and later integrate in $D=4-\epsilon$ dimensions. In this case, the result is also finite but different from the above one being equal to
\bea\label{CFJ2}
{\cal L}_{CFJ,2}=\frac{7e^2}{36\pi^2}\epsilon^{\lambda\mu\nu\rho}b_{\lambda}A_{\mu}\partial_{\nu}A_{\rho}.
\eea
Due to the presence of the $\gamma_5$ matrix in above calculations, we can also use 't Hooft-Veltman prescription \cite{tHooft}, where the $\gamma_5$ matrix is moved to the rightmost position, and only afterwards the dimensional regularization is performed. Then, we obtain the result 
\bea
{\cal L}_{CFJ,3}=-\frac{5e^2}{9\pi^2}\epsilon^{\lambda\mu\nu\rho}b_{\lambda}A_{\mu}\partial_{\nu}A_{\rho}.
\eea

Also, to simplify the expression (\ref{Pia}), we can employ the identity:
\bea
\gamma^{\mu}\gamma^{\lambda}\gamma^{\nu}=\eta^{\mu\lambda}\gamma^{\nu}+\eta^{\nu\lambda}\gamma^{\mu}-\eta^{\mu\nu}\gamma^{\lambda}-i\epsilon^{\sigma\mu\lambda\nu}\gamma_{\sigma}\gamma_5,
\eea
so that, we have
\bea
\label{eq13}
\Pi^{\lambda\tau}_a(p)&=&{\rm tr}\int\frac{d^4k}{(2\pi)^4}
\epsilon^{\gamma\lambda\mu\nu}\gamma_{\gamma}\gamma_5
\frac{\slashed{k}+M}{k^2-M^2}(\eta_{\nu\alpha}-\frac{1}{3}\gamma_{\nu}\gamma_{\alpha}-\frac{2k_{\nu}k_{\alpha}}{3M^2}-\frac{\gamma_{\nu}k_{\alpha}-\gamma_{\alpha}k_{\nu}}{3M})
\times\nonumber\\&\times&
\slashed{b}\gamma_5\eta^{\alpha\beta}\frac{\slashed{k}+M}{k^2-M^2}(\eta_{\beta\rho}-\frac{1}{3}\gamma_{\beta}\gamma_{\rho}-\frac{2k_{\rho}k_{\beta}}{3M^2}-\frac{\gamma_{\beta}k_{\rho}-\gamma_{\rho}k_{\beta}}{3M})\times\nonumber\\&\times&
\epsilon^{\delta\tau\rho\sigma}\gamma_{\delta}\gamma_5
\left(\frac{\slashed{k}+M}{k^2-M^2}\slashed{p}\frac{\slashed{k}+M}{k^2-M^2}(\eta_{\sigma\mu}-\frac{1}{3}\gamma_{\sigma}\gamma_{\mu}-\frac{2k_{\sigma}k_{\mu}}{3M^2}-\frac{\gamma_{\sigma}k_{\mu}-\gamma_{\mu}k_{\sigma}}{3M})+\right.\nonumber\\ &+&\left.
\frac{\slashed{k}+M}{k^2-M^2}
\left(\frac{2(k_{\sigma}p_{\mu}+k_{\mu}p_{\sigma})}{3M^2}+\frac{\gamma_{\sigma}p_{\mu}-\gamma_{\mu}p_{\sigma}}{3M}\right)
\right)+O(p^2).
\eea
Now, we calculate the trace straightforwardly without employing 't Hooft-Veltman prescription in $D$ dimensions. The corresponding contribution to the effective Lagrangian is
\bea
{\cal L}_{CFJ,4}=\frac{311e^2}{648\pi^2}\epsilon^{\lambda\mu\nu\rho}b_{\lambda}A_{\mu}\partial_{\nu}A_{\rho}.
\eea
Interestingly, if we calculate all traces in (\ref{eq13}) in $4$ dimensions and later integrate in $D$ dimensions as above, we obtain the same result (\ref{CFJ2}).

So, we have found that the result is finite within all these approaches but strongly depends on the regularization scheme. This is reasonable since, as we already noted, the contribution is badly divergent, and, as it is known, superficially divergent contributions can yield finite results only if they display some kind of the $\infty-\infty$ indeterminate form, so that all pole parts mutually cancel (this situation takes place, e.g., for calculation of CFJ and aether terms in some simplest LV extensions of QED, see discussion in \cite{ourLV} and references therein). 

The possible higher-derivative (Myers-Pospelov or higher-derivative CFJ) contributions can be calculated along the same lines as in the usual QED \cite{HDLV}. However, unlike that paper, we concentrate in terms of the first order in $b_{\mu}$ only, as it was done in \cite{MWS}, which is rather natural since the $b_{\mu}$ vector is expected to be small \cite{datatables}.

Within our paper, we calculate the purely minimal higher-derivative contribution. Explicitly, we consider the expressions (\ref{twomin}), where all orders in the external $p_{\mu}$ are taken into account:
\bea
\label{twominhigh}
\Pi_a&=&-e^2{\rm tr}\int\frac{d^4k}{(2\pi)^4}A_{\lambda}(-p)A_{\tau}(p)
(\gamma^{\lambda}\eta^{\mu\nu}-\gamma^{\mu}\eta^{\nu\lambda}-\gamma^{\nu}\eta^{\mu\lambda}+\gamma^{\mu}\gamma^{\lambda}\gamma^{\nu})
\times\nonumber\\&\times&
\frac{\slashed{k}+M}{k^2-M^2}(\eta_{\nu\alpha}-\frac{1}{3}\gamma_{\nu}\gamma_{\alpha}-\frac{2k_{\nu}k_{\alpha}}{3M^2}-\frac{\gamma_{\nu}k_{\alpha}-\gamma_{\alpha}k_{\nu}}{3M})
\slashed{b}\gamma_5\eta^{\alpha\beta}\times\nonumber\\&\times&
\frac{\slashed{k}+M}{k^2-M^2}(\eta_{\beta\rho}-\frac{1}{3}\gamma_{\beta}\gamma_{\rho}-\frac{2k_{\beta}k_{\rho}}{3M^2}-\frac{\gamma_{\beta}k_{\rho}-\gamma_{\rho}k_{\beta}}{3M})
\times\nonumber\\&\times&
(\gamma^{\tau}\eta^{\rho\sigma}-\gamma^{\rho}\eta^{\tau\sigma}-\gamma^{\sigma}\eta^{\tau\rho}+\gamma^{\rho}\gamma^{\tau}\gamma^{\sigma})
\times\nonumber\\&\times&
\frac{\slashed{k}+\slashed{p}+M}{(k+p)^2-M^2}(\eta_{\sigma\mu}-\frac{1}{3}\gamma_{\sigma}\gamma_{\mu}-\frac{2(k+p)_{\sigma}(k+p)_{\mu}}{3M^2}-\frac{\gamma_{\sigma}(k+p)_{\mu}-\gamma_{\mu}(k+p)_{\sigma}}{3M}),\nonumber\\
\Pi_b&=&-e^2{\rm tr}\int\frac{d^4k}{(2\pi)^4}A_{\lambda}(-p)A_{\tau}(p)
(\gamma^{\lambda}\eta^{\mu\nu}-\gamma^{\mu}\eta^{\nu\lambda}-\gamma^{\nu}\eta^{\mu\lambda}+\gamma^{\mu}\gamma^{\lambda}\gamma^{\nu})
\times\nonumber\\&\times&
\frac{\slashed{k}+M}{k^2-M^2}(\eta_{\nu\rho}-\frac{1}{3}\gamma_{\nu}\gamma_{\rho}-\frac{2k_{\nu}k_{\rho}}{3M^2}-\frac{\gamma_{\nu}k_{\rho}-\gamma_{\rho}k_{\nu}}{3M})
\times\nonumber\\&\times&
(\gamma^{\tau}\eta^{\rho\sigma}-\gamma^{\rho}\eta^{\tau\sigma}-\gamma^{\sigma}\eta^{\tau\rho}+\gamma^{\rho}\gamma^{\tau}\gamma^{\sigma})
\times\nonumber\\&\times&
\frac{\slashed{k}+\slashed{p}+M}{(k+p)^2-M^2}(\eta_{\sigma\alpha}-\frac{1}{3}\gamma_{\sigma}\gamma_{\alpha}-\frac{2(k+p)_{\sigma}(k+p)_{\alpha}}{3M^2}-\frac{\gamma_{\sigma}(k+p)_{\alpha}-\gamma_{\alpha}(k+p)_{\sigma}}{3M})
\slashed{b}\gamma_5\eta^{\alpha\beta}
\times\nonumber\\&\times&
\frac{\slashed{k}+\slashed{p}+M}{(k+p)^2-M^2}(\eta_{\beta\mu}-\frac{1}{3}\gamma_{\beta}\gamma_{\mu}-\frac{2(k+p)_{\beta}(k+p)_{\mu}}{3M^2}-\frac{\gamma_{\beta}(k+p)_{\mu}-\gamma_{\mu}(k+p)_{\beta}}{3M}).
\eea
It is clear, by symmetry reasons, that we, as above, can employ the definitions (\ref{defs}) for the self-energy tensors. Using again the 't Hooft-Veltman prescription and the Feynman parametrization, we obtain the following all-order result:
\bea
\Pi_a &=& -\left(\frac{20 p^2}{81 \pi ^2 \epsilon' M^2}-\frac{p^4}{27 \pi ^2 \epsilon' M^4}-\frac{p^6}{162 \pi ^2 \epsilon' M^6}\right)e^2\epsilon^{\lambda\mu\nu\rho}b_{\lambda}A_{\mu}(-p)p_{\nu}A_{\rho}(p)\nonumber\\
&+& \left(24 M^2 \sqrt{p^{10} \left(4 M^2-p^2\right)}+2 \sqrt{p^{14} \left(4 M^2-p^2\right)}-534 M^6 \sqrt{p^2 \left(4 M^2-p^2\right)}\right.\nonumber\\
&-& \left. 179 M^4 \sqrt{p^6 \left(4 M^2-p^2\right)}+6 \left(104 p^2 M^6-54 p^4 M^4+4 p^6 M^2+p^8-4 M^8\right)\right.\nonumber\\  
&\times& \left. \csc^{-1}\left(\frac{2M}{\sqrt{p^2}}\right)\right)\frac{e^2\epsilon^{\lambda\mu\nu\rho}b_{\lambda}A_{\mu}(-p)p_{\nu}A_{\rho}(p)}{972 \pi ^2 M^6 \sqrt{p^2 \left(4 M^2-p^2\right)}},
\eea
where $\frac1{\epsilon'}=\frac1{\epsilon}-\ln\frac{M}{\mu'}$, with $\epsilon=4-D$ and $\mu'^2=4\pi\mu^2e^{-\gamma}$. So, we see that the terms of third, fifth, and seventh orders in derivatives diverge.
By taking the limit $p^2 \ll M^2$ $(M\neq 0)$ in the finite contribution above, we obtain
\bea
\Pi_{a,fin} &=& -\frac{5e^2}{9\pi^2}\epsilon^{\lambda\mu\nu\rho}b_{\lambda}A_{\mu}(-p)p_{\nu}A_{\rho}(p) -\frac{2e^2p^2}{81\pi^2M^2}\epsilon^{\lambda\mu\nu\rho}b_{\lambda}A_{\mu}(-p)p_{\nu}A_{\rho}(p),
\eea
i.e., the CFJ term is confirmed to be finite unlike the higher-derivative contributions. Apparently this signalizes that higher-derivative terms do not contribute to the possible chiral anomaly whose arising is natural to expect in a whole analogy with the LV QED.

One more important term which can arise in the effective action is the lower CPT-even term, that is, the aether term. To simplify the calculations, we generate it not through introducing the $\slashed{b}\gamma_5$ term as in the previous examples, but with the use of the additional magnetic-like nonminimal coupling representing itself as a straightforward generalization of the magnetic coupling used in \cite{aether} and allowing to keep only zero order in derivatives from the contribution of the corresponding Feynman diagram:
\bea
V_g=g\bar{\psi}^{\alpha}\epsilon^{\mu\nu\rho\sigma}\gamma_{\mu}b_{\nu}F_{\rho\sigma}\psi_{\alpha}.
\eea
The relevant contribution looks like
\bea
\label{ae}
S_{ae}=-\frac{g^2}{2}\int \frac{d^4p}{(2\pi)^4}\epsilon^{\mu\nu\rho\sigma}b_{\nu}F_{\rho\sigma}(-p)\epsilon^{\mu'\nu'\rho'\sigma'}b_{\nu'}F_{\rho'\sigma'}(p){\rm tr}\int\frac{d^4k}{(2\pi)^4}\gamma_{\mu}G^{\alpha\alpha'}(k)\gamma_{\mu'}G_{\alpha'\alpha}(k).
\eea
It is given by the Feynman diagram:

\begin{figure}[htbp]
	\includegraphics{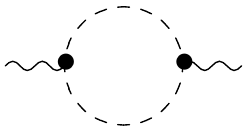}
	\caption{Aether-like contributions to the two-point function of the vector field.}
	\label{fig:2}       
\end{figure}

Explicitly, the contribution (\ref{ae}) is written as
\bea
\label{ae1}
S_{ae}&=&-\frac{g^2}{2}\int \frac{d^4p}{(2\pi)^4}\epsilon^{\mu\nu\rho\sigma}b_{\nu}F_{\rho\sigma}(-p)\epsilon^{\mu'\nu'\rho'\sigma'}b_{\nu'}F_{\rho'\sigma'}(p)\times\nonumber\\&\times&
{\rm tr}\int\frac{d^4k}{(2\pi)^4}\gamma_{\mu}
\frac{\slashed{k}+M}{k^2-M^2}(\eta^{\alpha\alpha'}-\frac{1}{3}\gamma^{\alpha}\gamma^{\alpha'}-\frac{2k^{\alpha}k^{\alpha'}}{3M^2}-\frac{\gamma^{\alpha}k^{\alpha'}-\gamma^{\alpha'}k^{\alpha}}{3M})\times\nonumber\\&\times&
\gamma_{\mu'}
\frac{\slashed{k}+M}{k^2-M^2}(\eta_{\alpha'\alpha}-\frac{1}{3}\gamma_{\alpha'}\gamma_{\alpha}-\frac{2k_{\alpha'}k_{\alpha}}{3M^2}-\frac{\gamma_{\alpha'}k_{\alpha}-\gamma_{\alpha}k_{\alpha'}}{3M}).
\eea
Calculating this expression straightforwardly with use of the dimensional regularization, we find
\bea
S_{ae}=\frac{g^2M^2}{6\pi^2}\left(\frac{1}{\epsilon'}+\frac{1}{12}\right)\int d^4x(4b_{\mu}F^{\mu\nu}b^{\lambda}F_{\lambda\nu} -2b^2F_{\mu\nu}F^{\mu\nu}).
\eea
So, we generated the aether term, but unlike the usual case \cite{aether}, it displays the divergence. We note that, in principle, the presence of this divergence is rather natural since our model is worse from the renormalization viewpoint than the theory considered in \cite{aether}. 

At the same time, it is important to notice that the aether term is proportional to the second order in the LV vector $b_{\mu}$. Therefore, it is highly suppressed. The same conclusion is valid if we try to generate the aether term on the base of our initial Lagrangian (\ref{lagr}) without employing nonminimal vertices. So, if we start from any CPT-odd gauge-RS coupling, the aether term must be of the second order in LV parameters. The next nontrivial contribution of the first order in the $b_{\mu}$ will involve higher derivatives.

Let us discuss our results. We have proved that, despite a horrible sextic divergence (to the best of our knowledge, such a divergence never was reported, perhaps except for nonlocal theories whose Lagrangian involves $\Box^{-1}$ or $\Box^{-2}$ operators, as occurs in some nonlocal gravity models, see, e.g., \cite{Mitra}), the result for the CFJ term is finite but strongly dependent on the regularization scheme, i.e., ambiguous. In a whole analogy with the usual LV QED, it is natural to expect that this ambiguity can be related to some analogue of the Adler-Bell-Jackiw (ABJ) anomaly (for the usual QED, this relation has been discussed in \cite{JackAmb}). The gauge covariance reasons allow us to conclude that this result can be straightforwardly generalized to a non-Abelian case, resulting in non-Abelian CFJ and aether terms. We note that our results do not depend on the mass parameter $M$, so it is natural to conclude that even coupling of the massless (and hence gauge invariant) RS field to the electromagnetic field will yield the CFJ term. From another side, the higher-derivative CFJ-like terms, involving up to seven derivatives, diverge. Comparing this situation with the usual LV QED where the three-derivative terms were shown to be finite, and, in certain cases, ambiguous \cite{HDLV}, and taking into account the relation between ambiguities and anomalies claimed in \cite{JackAmb}, we can conclude that the possible analogue of the chiral anomaly in the case of the presence of RS fields will not include higher derivatives.

This study can be generalized to the case of the RS field coupled to the gravity, where it is reasonable to expect arising of the $4D$ gravitational Chern-Simons term. We plan to discuss the anomalies in the theory of the Rarita-Schwinger field coupled to a gauge field and to gravity in a forthcoming paper.

{\bf Acknowledgments.} This work was partially supported by Conselho Nacional de De- senvolvimento Cient\'ifico e Tecnol\'ogico (CNPq). The work of A. Yu. P. has been partially supported by the CNPq project No. 301562/2019-9.

\end{document}